%% file: main.tex
\documentclass[10pt,journal,compsoc]{IEEEtran}
\ifCLASSOPTIONcompsoc
  \usepackage[nocompress]{cite}
\else
  \usepackage{cite}
\fi
\ifCLASSINFOpdf
   \usepackage[pdftex]{graphicx}
\else
\fi
\usepackage{amsmath}
\usepackage{hyperref}

\ifCLASSOPTIONcompsoc
 \usepackage[caption=false,font=footnotesize,labelfont=sf,textfont=sf]{subfig}
\else
 \usepackage[caption=false,font=footnotesize]{subfig}
\fi

\usepackage{xcolor}
\usepackage{wrapfig}

\begin{document}

\newcommand{\sysname}{Outcome-Explorer}
%
\title{\sysname{}: A Causality Guided Interactive Visual Interface for Interpretable Algorithmic Decision Making}

%
%
%
%

\author{Md Naimul Hoque
        and Klaus Mueller,~\IEEEmembership{Senior Member, IEEE}
\IEEEcompsocitemizethanks{\IEEEcompsocthanksitem  Md Naimul Hoque was with the Computer Science Department, Stony Brook University. He is now with the College of Information Studies, University of Maryland, College Park, MD 20742 USA.
E-mail: nhoque@umd.edu}
\IEEEcompsocitemizethanks{\IEEEcompsocthanksitem 
Klaus Mueller is with the Computer Science Department, Stony Brook University, Stony Brook, NY 11794\protect\\
E-mail: mueller@cs.stonybrook.edu}
\thanks{Manuscript received January 4, 2021; revised xxx}}

%
%

\markboth{IEEE Transaction on Visualization and Computer Graphics, 2021. Author's Version. \url{https://doi.org/10.1109/tvcg.2021.3102051}}%
{Shell \MakeLowercase{\textit{et al.}}: Bare Demo of IEEEtran.cls for Computer Society Journals}
%



\IEEEtitleabstractindextext{%
\begin{abstract}
The widespread adoption of algorithmic decision-making systems has brought about the necessity to interpret the reasoning behind these decisions. The majority of these systems are complex black box models, and auxiliary models are often used to approximate and then explain their behavior. However, recent research suggests that such explanations are not overly accessible to lay users with no specific expertise in machine learning and this can lead to an incorrect interpretation of the underlying model. In this paper, we show that a predictive and interactive model based on causality is inherently interpretable, does not require any auxiliary model, and allows both expert and non-expert users to understand the model comprehensively. To demonstrate our method we developed Outcome Explorer, a causality guided interactive interface, and evaluated it by conducting think-aloud sessions with three expert users and a user study with 18 non-expert users. All three expert users found our tool to be comprehensive in supporting their explanation needs while the non-expert users were able to understand the inner workings of a model easily.



\end{abstract}

\begin{IEEEkeywords}
Explainable AI, Causality, Visual Analytics, Human-Computer Interaction.
\end{IEEEkeywords}}

\maketitle

\IEEEdisplaynontitleabstractindextext

%
\IEEEpeerreviewmaketitle
\input{sections/1-introduction.tex}

\input{sections/2-related_works}
\input{sections/3-background}

\input{sections/4-design.tex}

\input{sections/6-implementation.tex}
\input{sections/5-use_cases}

\input{sections/7-evaluation}
\input{sections/8-discussion_conclusion}

\ifCLASSOPTIONcompsoc
  \section*{Acknowledgments}
\else
  \section*{Acknowledgment}
\fi
This research was partially supported by NSF grants IIS 1527200 and 1941613.



\ifCLASSOPTIONcaptionsoff
  \newpage
\fi



\bibliographystyle{IEEEtran}
\bibliography{references.bib}
%



%


\begin{IEEEbiography}
[{\includegraphics[width=1in,height=1.25in,clip,keepaspectratio]{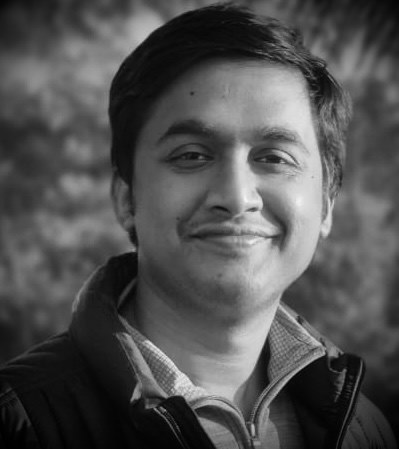}}]
{Md Naimul Hoque} is currently a PhD student at the College of Information Studies, University of Maryland, College Park. Previously, he obtained an  M.S. in Computer Science degree from Stony Brook University. His current research interests include explainable AI, visual analytics, and human-computer interaction. For more information, see https://naimulh0que.github.io
\end{IEEEbiography}
\vspace{-200pt}
\begin{IEEEbiography}[{\includegraphics[width=1in,height=1.25in,clip,keepaspectratio]{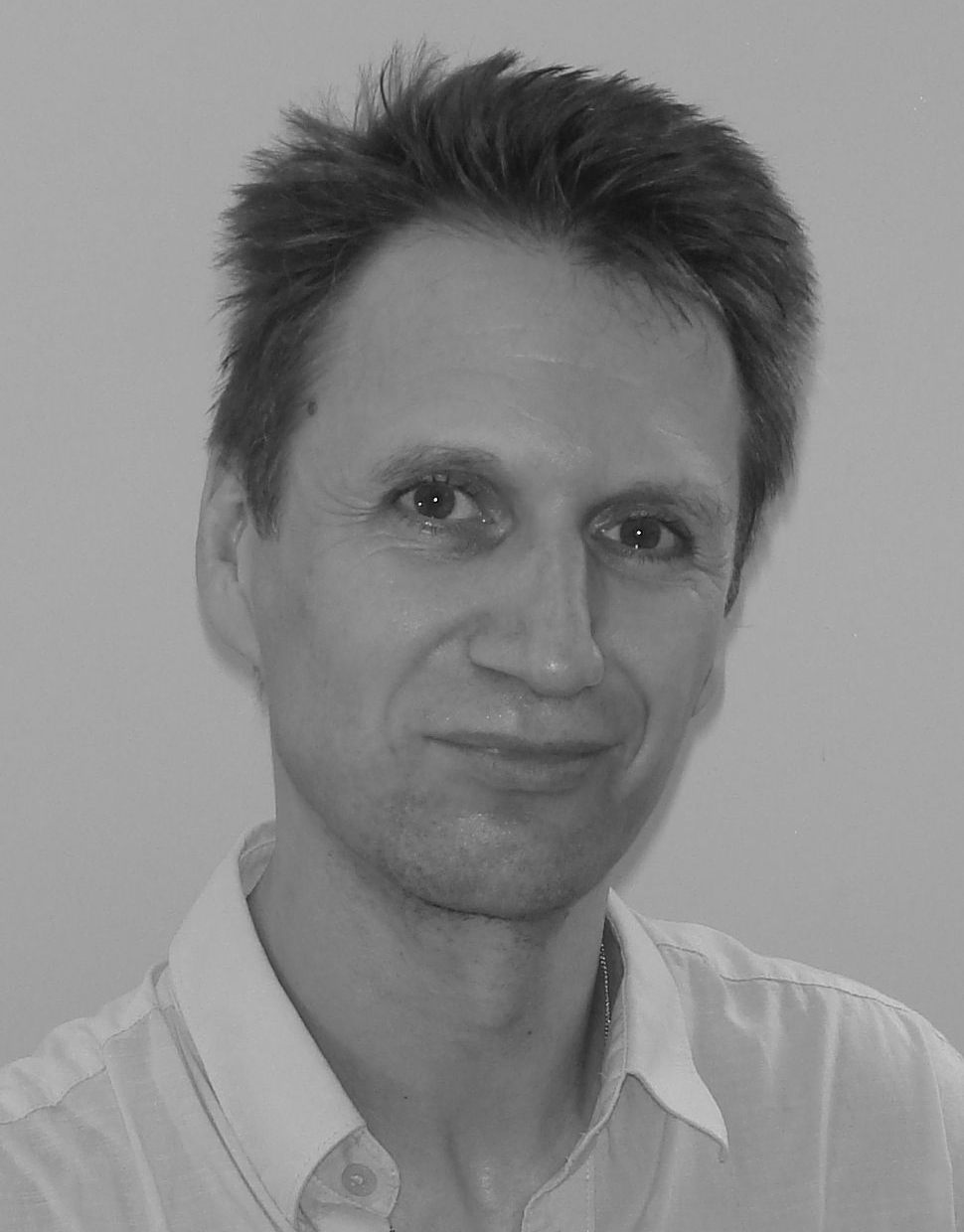}}]{Klaus Mueller}
has a
PhD in computer science 
and is currently a professor
of computer science at Stony Brook University
and is a senior scientist at Brookhaven
National Lab. His current research interests
include explainable AI, visual analytics, data science, and medical imaging. He won the US
National Science Foundation Early CAREER Award, the SUNY Chancellor’s
Award for Excellence in Scholarship \& Creative Activity, and the IEEE CS Meritorious
Service Certificate. His 200+ papers
were cited over 10,000 times. For more information, see
http://www.cs.sunysb.edu/ mueller
\end{IEEEbiography}







\end{document}

%% file: sections/1-introduction.tex
\IEEEraisesectionheading{\section{Introduction}\label{sec:introduction}}

\IEEEPARstart{I}{n} recent years, algorithmic and automated decision-making systems have been deployed in many critical application areas across society~\cite{brennan2009evaluating,chouldechova2018case,obermeyer2019dissecting}. It has been shown, however, that these systems can exhibit discriminatory and biased behaviors (e.g.,~\cite{angwin2016machine,buolamwini2018gender,keyes2018misgendering}). It is thus imperative to create mechanisms by which humans can interpret and investigate these automated decision-making processes.

Several algorithmic~\cite{NIPS2017_7062,ribeiro2016should} and visual analytics~\cite{hohman2018visual,abdul2018trends, amershi2015modeltracker, kahng2017cti, hohman2019gamut, hohman2019s} solutions have been proposed to address this need. However, a shortcoming of these systems is that they have been predominantly developed from a model-builders perspective and as such do not support common lay (\textit{non-expert}) users with no specific expertise in machine learning\cite{cheng2019explaining,miller2019explanation}. It is these individuals, however, who are typically the recipients of a decision algorithm's outcome~\cite{cheng2019explaining,miller2019explanation}. 

This shortcoming is directly addressed in the 2016 EU General Data Protection Regulation (GDPR) which mandates that non-expert users who are directly impacted by algorithmic decisions have a ''right to explanation". 
According to the GDPR users should receive these explanations in a \emph{concise, transparent, intelligible, and easily accessible form}~\cite{voigt2017eu}. The GDPR has since become a model for laws beyond the EU, for example, the 2018 California Consumer Privacy Act (CCPA) bears many similarities with the GDPR. In addition, several recent studies~\cite{cheng2019explaining, shen2020designing, miller2019explanation} have further confirmed the needs of non-expert users to interpret and understand machine-generated decisions.




 However, supporting non-expert users in a explainable AI (XAI) platform is challenging since they have different goals, reasons, and skill sets for interpreting a machine learning model than expert users. A model-builder such as a machine learning practitioner with significant data science expertise
 will want to interpret a model to ensure its accuracy and fairness. A non-expert user, on the other hand, will want to interpret the model to understand the service the model facilitates and delivers, and gain trust into it. In most cases, these users will not have background in data science and machine learning and so require an easy-to-understand visual representation of the model.
 


To exhibit human-friendly interpretability, a predictive model also needs to produce answers to explanation queries such as ``Why does this model make such decisions?'', ``What if I change a particular input feature?'' or ``How will my action change the decision?''~\cite{moraffah2020causal}. Existing methods and systems often employ an auxiliary model (post-hoc) to first approximate the original black-box model and then answer such questions since black-box models do not readily provide these explanations ~\cite{NIPS2017_7062,ribeiro2016should,ming2018rulematrix}. However, recent research suggests that such approximations can lead to incorrect interpretations of the underlying model and further complicate model understanding~\cite{rudin2019stop,kumar2020problems}. 
Thus, an XAI interface for non-expert users should be inherently interpretable (i.e. directly observable)~\cite{rudin2019stop} and be able to answer causal questions without 
requiring an auxiliary model.

In this paper, we show that a predictive model based on causality (i.e., a causal model) meets the aforementioned criteria for a model that is understandable by non-expert users. The inner-workings of a causal model is directly observable through a Directed Acyclic Graph (DAG), making it inherently interpretable. The causal DAG is an intuitive representation and based on it and through interactions defined on it, a user can gain a good understanding of how variables are related to each other and how they affect the outcome variable, without the need for machine learning expertise. 
Further, causal models can provide truthful answers to causal explanation queries
without auxiliary models.

To support our proposed method, we first developed a computational pipeline that would facilitate predictions in a causal model. This was needed since causal models do not support predictions by default. Informed by prior research and a formative study with ten non-expert users, we designed and developed \sysname{}, an interactive algorithmic decision-making interface based on causality. \sysname{} lets both non-expert and expert users interact with the causal DAG and supports common XAI functionalities such as answering What-If questions, exploring nearest neighbors, and comparing data instances. To evaluate the effectiveness of \sysname{}, we first invited three expert users (machine learning researchers) to develop a causal model and then interpret the model using our tool. All three expert users found \sysname{} 
comprehensive in terms of both its causal and its explanation functionalities.

We then conducted a user study with $18$ non-expert lay users. We sought to understand how \sysname{} would help non-expert users in interpreting a predictive model, in comparison to a  popular post-hoc XAI method, SHAP~\cite{NIPS2017_7062}. The study revealed that participants were able to reduce interactions with variables that did not affect the outcome by $47\%$, meaning that participants understood which variables to change while using our tool. A similar reduction ($36\%$) was also found for the \textit{magnitude} of changes made to non-impacting variables. Hence, the interactions of the participants became more efficient which aided their understanding of the model.  These outcomes confirm the high potential of \sysname{} in both XAI research and application.


Our research contributions are as follows; we describe:
\begin{itemize}
\vspace{-0.2cm}
    \item A mechanism that allows an interactive assessment of prediction accuracy in a causal model.
    \item The design and implementation 
    of a causality-guided interactive visual tool, \sysname{}, to support the model explanation needs of both expert and non-expert users.
    \item Results from think-aloud sessions with three expert users which revealed that experts are able to build and interpret a predictive causal model correctly.
    \item A user study with $18$ non-expert users
which revealed that our tool can help non-experts  
understand a predictive causal model and enable them to identify the input features relevant for prediction.
\end{itemize}

Our paper is organized as follows. Section 2 and 3 describe related work and background. Section 4 presents our formative user study with non-expert users. Section 5 presents our design guidelines. Section 6 describes our interactive visual interface. Section 7 demos a use case. Section 8 presents the outcome of our system evaluation. Sections 9 and 10 present a discussion and conclusions. 

 \begin{figure*}
    \centering
    \includegraphics[width=0.8\textwidth]{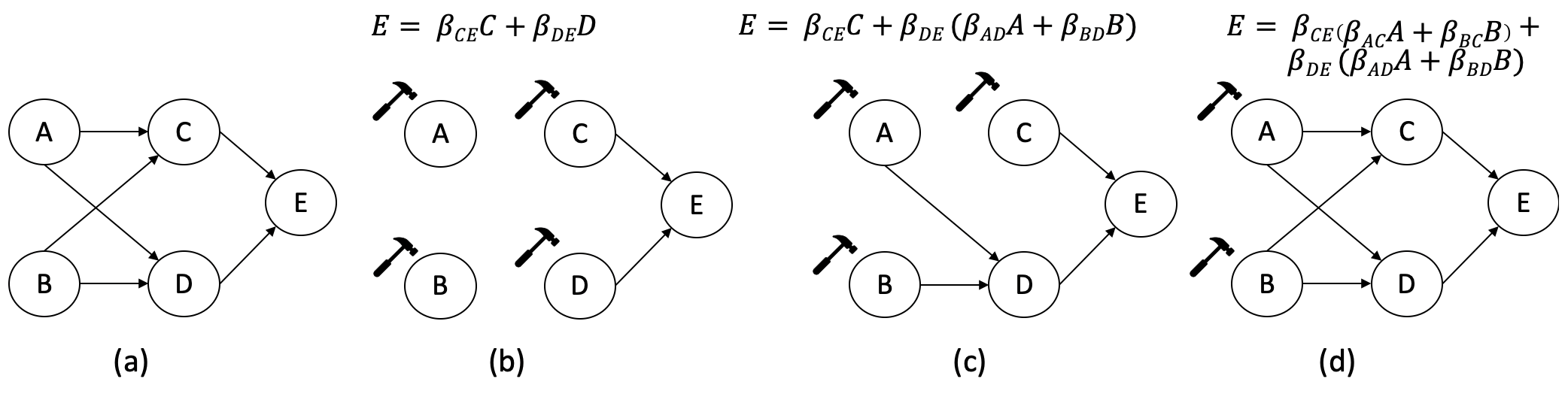}
    \caption{Prediction in a causal model. The hammer icon represents intervention. (a) true causal model. (b) Interventions on all feature variables. The causal links leading to node C and D are removed since the values of C and D are set externally. (c) Interventions on node A, B, and C. (d) interventions on node A and B. In the path equations above models (b)-(d), the $\beta$ are  standardized regression coefficients estimated from the data. }
    \label{fig:path_diagram}
\end{figure*}

%% file: sections/2-related_works.tex
\section{Related Work}

 Current tools and methods in XAI can be broadly categorized into two distinct categories: (1) post-hoc explanation, and (2) explanation via interpretability. Post-hoc explanation methods explain the prediction of a model using local estimation techniques, without showing or explaining the workflow of the model. Instead, they give very detailed information on the impact of the model variables at a user-chosen data instance and seek to build trust into the model one instance at a time. Prominent examples of post-hoc explanation methods are SHAP \cite{NIPS2017_7062} and LIME \cite{ribeiro2016should}. They are model agnostic and can explain the predictions of any machine learning model. While post-hoc methods are shown to be effective for explaining the decisions of complex black-box models, recent research has argued that these explanations can be misleading, and can complicate model understanding even more ~\cite{rudin2019stop,kumar2020problems}. 



To address these shortcomings, there has been a growing interest in devising models that are inherently interpretable.
The Generalized Additive Model (GAM)~\cite{hastie1990generalized} and the Bayesian Case Model (BCM)~\cite{kim2014bayesian} are examples of this kind of model where the workflow is directly observable by a human user. In this paper, we advocate for a causal model since it has a natural propensity towards interpretability. According to Pearl~\cite{pearl2018book}, the aim of causal models is to ``\textit{explain to us why things happened, and why they
responded the way they did}'' which aligns perfectly with the objective of XAI. 
\textcolor{black}{They have found frequent use to support reasoning about different forms of bias and discrimination \cite{glymour2019measuring, wu2019pc, zhang2018fairness, madras2019fairness, loftus2018causal, kusner2018causal, khademi2019fairness} but none uses interactive visualization within the model as an XAI paradigm. \sysname{} aims at bridging XAI, causal modeling, and visual analytics.  }

\subsection{Visual Analytics and XAI}
People are more likely to understand a system if they can tinker and interact with it \cite{dietvorst2018overcoming}. In that spirit, interactive visual systems have proven to be an effective way to understand algorithmic decision-making. These interactive systems broadly categorize into the aforementioned post-hoc explanation and interpretable interfaces. Examples of interactive post-hoc explanation interfaces include the What-if Tool \cite{wexler2019if}, RuleMatrix \cite{ming2018rulematrix}, Vice \cite{gomez2020vice}, Model Tracker \cite{amershi2015modeltracker}, Prospector \cite{krause2016interacting}, and others. These tools use scatter plots, line plots, and text interfaces to allow users to query and compare the outcomes of different decision models, but without showing the model itself. Though this helps to understand the model's behavior in a counterfactual sense (the 'if'), it does not explain, and allow a user to play with the reasoning flow within the system (the 'why'). 

On the other hand, interpretable interfaces such as GAMUT \cite{hohman2019gamut}, and SUMMIT \cite{hohman2019s} allow a user to interact with the model itself and provide explanations that are faithful to the model. While GAMUT is more conceptual in nature, pointing out the features an interactive interface should have to support model interpretation, SUMMIT specifically focuses on deep neural networks and the classification of images. It allows users to recognize how different classes of images are evaluated at different stages of the network and what features they have in common. Our tool, \sysname{}, also falls into the category of these systems, but focuses on causal networks and quantitative data. It specifically targets non-expert users and actively supports four of the six interface features specified in GAMUT. We chose this subset since they appear most suited for non-expert users.  





\subsection{Interactive Causal Analysis}

Several interactive systems have been proposed for visual causal analysis but none have been designed for algorithmic decision-making.
Wang and Mueller~\cite{wang2015visual, wang2017visual} proposed a causal DAG-based-modeler equipped with several statistical methods. The aspect of model understanding and what-if experience it provides to users is via observing how editing the causal network's edges affects the model’s quality via the Bayesian Information Criterion (BIC) score. Our tool, \sysname{}, on the other hand conveys the what-if experience by allowing users to change the values of the network nodes (the model variables) and observe the effect this has on the outcome and other variables.  At first glance both help in model understanding, but only the second is an experimenter’s procedure. It probes a process with different inputs and collects outcomes (predictions), using the causal edges to see the relations with ease. This kind of what-if analysis has appeared in numerous XAI interfaces~\cite{wexler2019if,hohman2019gamut,gomez2020vice} and we have designed ours specifically for causal models. The mechanism also appeals to self-determination and gamification which both play a key part in education and learning \cite{ryan2008self}. One might say that achieving a lower BIC score is also gamification, but a BIC score does not emotionally connect a person to the extent that a lower house price or a college admission does. It provides a storyline, fun and realism, all elements of  gamification~\cite{schell2008art}.


 Yan et al.~\cite{yan2020silva} proposed a method that allows users to edit a post-hoc causal model in order to reveal issues with algorithmic fairness. Their focus is primarily on advising analysts which causal relations to keep or omit to gain a fairer adjunct ML model. We, on the other hand, use the causal model itself for prediction and focus on supporting the \textit{right to explanation} of both expert and non-expert users.

The work by Xie et al.~\cite{xie2020visual} is closest to ours but it uses different mechanisms for value propagation and it also addresses a different user audience. 
They look at the problem in terms of distributions which essentially is a manager’s cohort perspective, while we look at specific outcomes from an individual’s perspective. Both are forms of what-if analyses but we believe the latter is more accessible to a person directly affected by the modeled process, and so is more amenable to non-expert users who do not think in terms of distributions, uncertainties, and probabilities. 

Finally, several visual analytics systems have been proposed for causal analysis in ecological settings. 
 CausalNet~\cite{onoue2018development} allows users to interactively explore and verify phenotypic character networks. A user can search for a subgraph and verify it using several statistical methods (e.g., Pearson Correlation and SEM) and textual descriptions of the relations, mined from a literature database. Natsukawa et al.~\cite{natsukawa2020visual} proposed a dynamic causal network based system to analyze evolving relationships between time dependant ecological variables. While these systems have been shown to be effective in analyzing complex ecological relations, they do not provide facilities for network editing and interventions which ours does. They also do not support what-if and counterfactual analyses which are instrumental for an XAI platform~\cite{wexler2019if,hohman2019gamut}.



In summary, the fundamental differences of
ours to the existing systems are: (1) introduction of a prediction mechanism; (2) allowing users to change variables  (intervention) by directly interacting with the nodes in the causal DAG; (3) visualizing the interplay between variables when applying/removing interventions; (4) supporting explanation queries such as what-if analyses, neighborhood exploration, and instance comparisons; and
(5) including non-expert users in the design process, supporting their explanation needs.

%% file: sections/3-background.tex
\section{Background}

We follow Pearl's Structural Causal Model (SCM) \cite{pearl2009causality} to define causal relationships between variables. According to SCM, causal relations between variables are expressed in a DAG, also known as Path Diagram. In a path diagram, variables are categorized as either exogenous ($U$) or endogenous ($V$). Exogenous variables have no parents in a path diagram and are considered to be independent and unexplained by the model. On the other hand, endogenous variables are fully explained by the model and presented as the causal effects of the exogenous variables. Figure \ref{fig:path_diagram} presents two exogenous variables ($A$ and $B$) and three endogenous variables ($C, D,$ and $E$). Formally, the Causal Model is a set of triples $(U, V, F)$ such that

\begin{itemize}
    \item U is the set of exogenous variables, and V is the set of endogenous variables.
    \item Structural equations~\cite{bentler1980linear} (F) is a set of functions $\{f_1, . . . , f_n\}$, one for each $V_i \subseteq V$, such that $V_i = f_i(pa_i
, U_{pa_i})$, $pa_i \subseteq V \setminus \{V_i\}$ and $U_{pa_i} \subseteq U$. 
\end{itemize}

The notation ``$pa_i$'' refers to the ``parents'' of $V_i$. 



\subsection{Causal Structure Search}
\label{causal_search}
The causal structure between variables (F) can be obtained in three different ways: (1) causal structure defined from domain-expertise or prior knowledge; (2) causal structure learned from automated algorithms; and (3) causal structure learned from mixed-initiative human-AI collaboration. 
 
 The first method is the prevalent way to operate causal analysis in domains such as social science or medical science where expert-users or researchers are solely responsible for defining the causal structure~\cite{hoyle1995structural}. In such scenarios, researchers utilize prior knowledge, domain expertise, and empirical evidence gathered from experiments such as randomized trials to hypothesized causal relations and then test the validity of the model through Structural Equation Modelling (SEM). Software such as ``IBM SPSS AMOS'', and ``Lavaan'' are build upon this principle.
 
 On the other hand (in the second approach), automated causal search algorithms utilize conditional independence tests to find causal structure among data~\cite{glymour2019review}. These algorithms help the user identify underlying causal structures between a large set of variables. The ``Tetrad'' software provides a comprehensive list of such algorithms.
 
The third approach combines the first two approaches. One pitfall of the automated algorithms is that they may not find the true causal structure since multiple causal structures can meet the constraints set out by the algorithms. In such scenarios, human verification is necessary to validate the causal structure obtained from automated algorithms. Prior research and empirical evidence suggest this human-centered research approach can identify the causal structure better than automated algorithms alone~\cite{shen2020challenges,wang2015visual,wang2017visual}.
 
 Finally, once the causal structure (F) is learned via any of the above approaches, we can use SEM to parameterize the model, obtaining the path/beta coefficients.

\subsection{Prediction}

 We define the outcome variable $Y$ as an endogenous variable, the variable we want to predict from a set of feature variables $X$.  In a causal model ($M$), estimating $Y$ from $X$ is analogous to applying an intervention (fixing variables to specific values) on $X$~\cite{tople2019alleviating}. This is achieved through the $do$ operator which simulates a causal intervention by deleting certain edges from $M$ while fixing $X$ to specific values~\cite{pearl2009causality}. The resultant causal model $M'$ is a subgraph of the original model $M$. Figure~\ref{fig:path_diagram}(b) shows $M'$ when intervention is applied to all variables of $X$. Note that the edges leading to Node $C$ and $D$ from the original model are removed in figure~\ref{fig:path_diagram}(b). This is because nodes $A$ and $B$ can no longer causally effect $C$ and $D$, once we fix them to specific values. In this scenario, $Y(E)$ can be estimated using this equation:
\begin{equation}
P_M(Y|do(X=x)) = P_{M'}(Y) = \beta_{CE} C+ \beta_{DE} D
\end{equation}
The $\beta$ are standardized regression coefficients estimated from SEM. Figure~\ref{fig:path_diagram}(c) and (d) present different intervention scenarios on the original model $M$ where the predictions $P_M$ follow the equations shown above the respective model. The key idea is that $Y$ is independent of its ancestors conditioned on all of its parents. Once we know the direct parents $Y$, other variables in the causal DAG can no longer influence $Y$.

%% file: sections/4-design.tex
\begin{figure}[t]
    \centering
    \includegraphics[width=0.9\columnwidth]{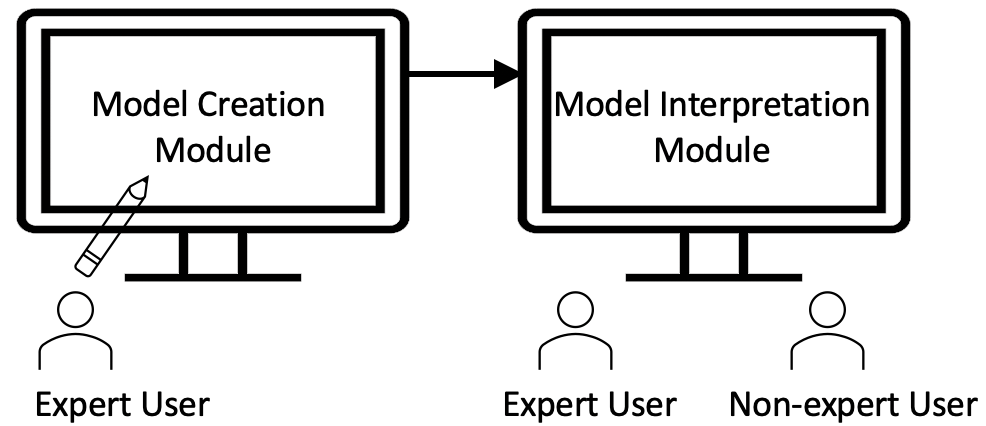}
    \caption{Two module design of \sysname{} with respective target users. An expert user would use the Model Creation module to create the causal model. After that, both expert and non-expert users would use the Model Interpretation module to interpret the causal model.}
    \label{fig:user_flow}
\end{figure}

\begin{figure*}
    \centering
    \includegraphics[width=0.95\textwidth]{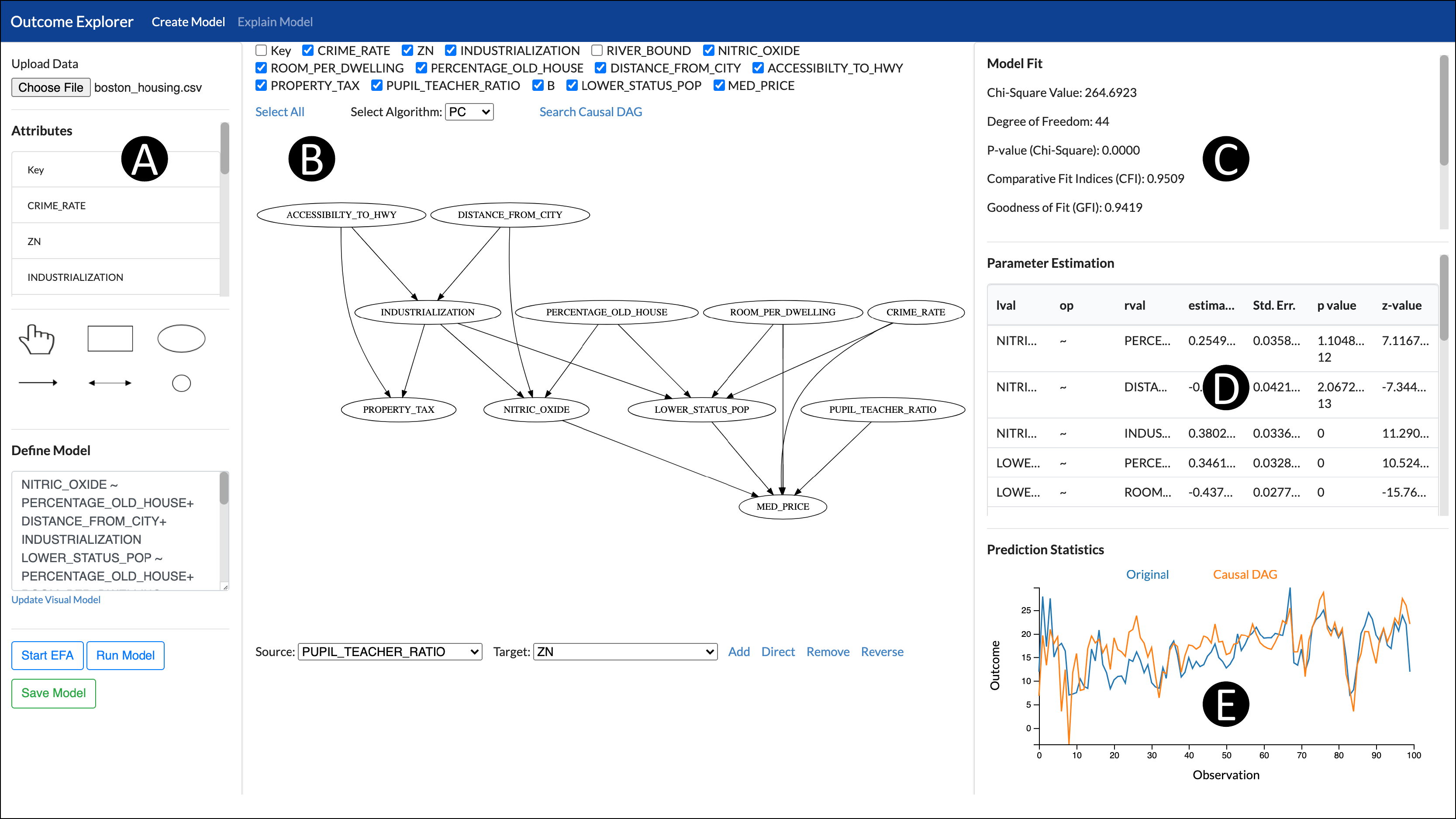}
    \caption{Model Creation Module of \sysname{}. A) Control Panel. B) Causal structure obtained from the search algorithms. Users can interactively add, remove, and direct edges in the causal structure. C) Model fit measures obtained from Structural Equation Modelling (SEM). D) Parameter estimation for each relations (beta coefficients). E) A line chart showing the prediction accuracy of the Causal Model on the test set. }
    \label{fig:create_model}
\end{figure*}

\section{Formative Study with Non-expert users}




We conducted formative interviews with 10 users (5 female, 5 male) to understand the expectations that non-expert users have of a decision-making interface. We recruited the participants (\textbf{P1-P10}) through social media posts and local mailing lists. Our inclusion criteria included familiarity with online decision-making services such as credit card approval and insurance services; algorithmic expertise was not required. We prompted the participants about their experience with these decision-making services. Their feedback is summarized in the following. 

\textbf{Need for transparency.}
  We noticed a general need for transparency among non-expert users. They appeared to prefer an automated system over human assistance, but feared the systems might not give them the optimal service. Several participants mentioned that automated platforms allowed them to obtain service quickly and efficiently, whereas to get human assistance they often had to wait on the phone for a long time (\textbf{P1-4, P6, P9}). Yet, several participants mentioned that eventually they needed to contact a human agent since specific rules and provisions were often not readily available in the automated systems (\textbf{P1-4, P6, P9}). Thus,  \sysname{} should be completely transparent and provide necessary explanations for decisions.
 
 
%


\textbf{How can I improve the decision?}
When asked about the process of evaluating algorithmic decisions, the participants mentioned that they would repeatedly update the input features to change the decision in their favor~(\textbf{P1-7, P9}). They would try to make sense of the underlying algorithm by changing the values of variables and then observe the effect this had on the outcome variable~(\textbf{P1-7, P9}). We note that this process is the same as obtaining~\textbf{local instance explanation} and  asking~\textbf{what-if (counterfactual) questions},  which are capabilities C1 and C3 identified by Hohman et al.~\cite{hohman2019gamut} as needed by expert users to interpret a single decision. Our tool should support C1 and C3 for non-expert users as well.


%

\textbf{How am I different than my friend?} Users of automated decision-making systems often employ a mental process of comparing themselves with others and try to make sense of why different people received different decisions. The participants shared several such cases where they wondered why they received one decision, while their friends received different decisions~(\textbf{P1-5, P7-8, P10}). We note that this is similar to \textbf{instance comparison} and \textbf{neighborhood exploration}, which are capabilities C2 and C4 identified by Hohman et al.~\cite{hohman2019gamut} that expert users should have to compare a data point to its nearest neighbors. Our study shows that our tool should support these capabilities also for non-expert users. 


\section{Design Guidelines}

Based on the insights gathered from our formative study, we formulate the following design guidelines:

\textbf{DG1. Supporting experts and non-experts via a two module design:}
The formative study revealed overlapping interests between expert and non-expert users to interpret predictive models. However, a model needs to be created before it can be interpreted. XAI interfaces typically accept trained models for this purpose~\cite{hohman2019gamut,hohman2019s,ming2018rulematrix}. However, at the time of the development of this work, no open-source software or package was available for human-centered causal analysis (the third method from Section~\ref{causal_search}).  Additionally, none of the existing tools supported prediction in a causal model. Hence, we decided to also support the creation of a predictive causal model. 

The methods described for creating a causal model in Section 3 requires substantial algorithmic and statistical expertise which can only be expected from an expert user. The relationship between expert and non-expert users follows the \emph{producer-consumer} analogy where an expert user will create and interpret the model for accurate and fair modeling, while a non-expert user will interpret this verified model to understand the service the model facilitates. To support this relationship, we decided that \sysname{} should have two different modules: \textbf{(1) Model Creation} module, and \textbf{(2) Model Interpretation} module. Figure~\ref{fig:user_flow} shows the two modules and their respective target users.



 
 



\textbf{DG2. Creating the Model:}
Using the Model Creation module, an expert user should be able to create a causal model interactively with the help of state-of-the-art techniques and evaluate the performance of the model.




\textbf{DG3. Interpreting the Causal DAG:}
The causal DAG is central to understanding a causal model. The visualization and interaction designed for the causal DAG in the Model Interpretation module should allow both expert and non-expert users to interpret the model correctly. Users should be able to set values to the input features in the DAG to observe the changes in the outcome. 





\textbf{DG4. Supporting Explanation Queries:}
The formative study revealed that non-expert users ask explanation queries (\textbf{C1-C4}) similar to those already well-studied in XAI research~\cite{hohman2019gamut}. Our tool should support these queries and they should be implemented keeping in mind the algorithmic and visualization literacy gap between expert and non-expert users.
 

\begin{figure*}
    \centering
    \includegraphics[width=0.95\textwidth]{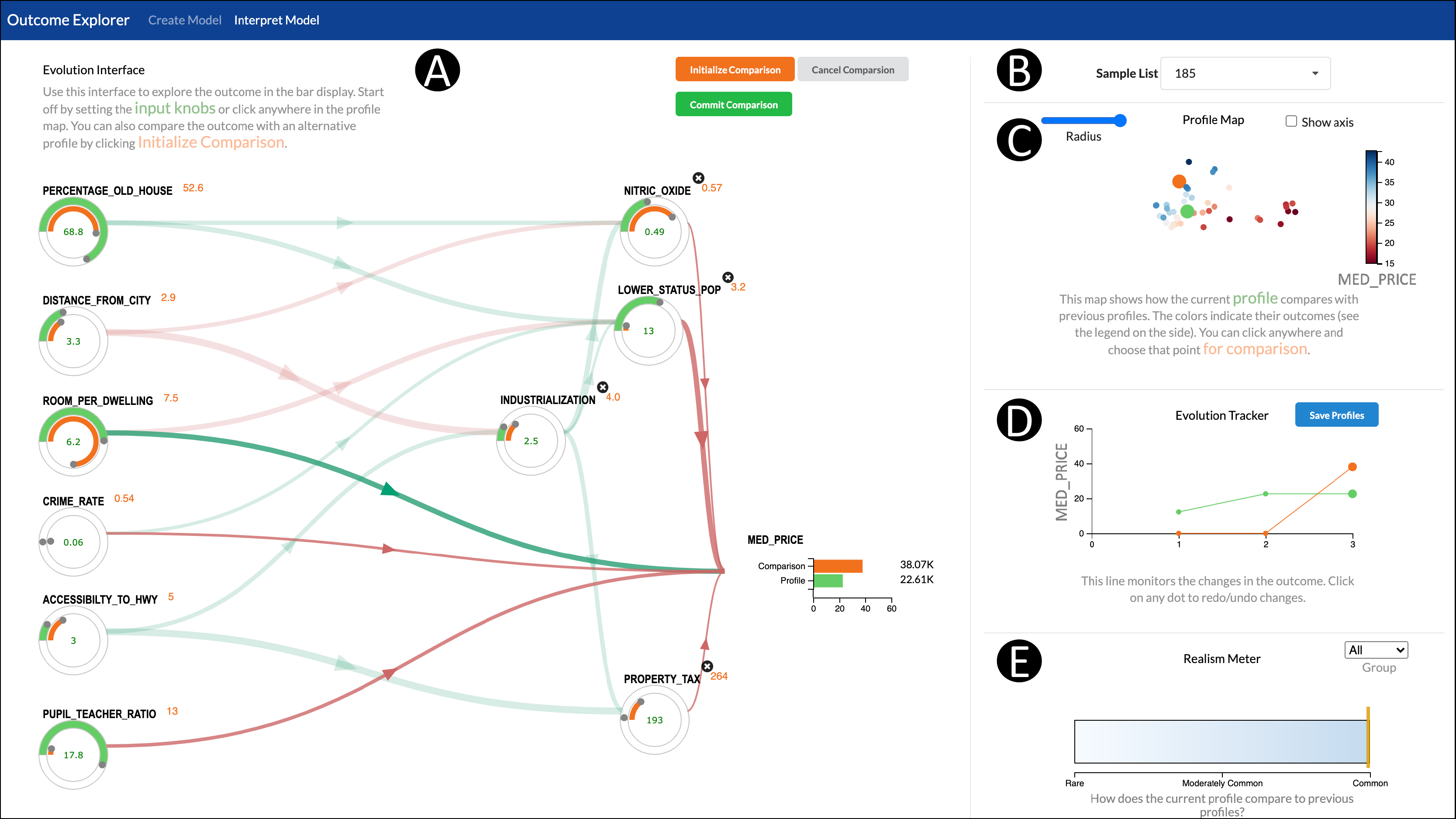}
    \caption{Model Interpretation Module of \sysname{}. A) Interactive causal DAG showing causal relations between variables. Each node includes two circular knobs (green and orange) to facilitate profile comparisons. The edge thickness and color depict the effect size and type of each edge. B) Sample selection panel. C) A biplot showing the position of green and orange profiles compared to nearest neighbors. D) A line chart to track the model outcome and to go back and forth between feature configuration. E) Realism meter allowing users to determine how common a profile is compared to other samples in the dataset. }
    \label{fig:explain_model}
\end{figure*}

\textbf{DG5. Input Feature Configuration:} Our tool is completely transparent and a non-expert user can change the input features freely in the interface. However, when engaging in this activity, it is possible that to obtain a certain outcome a user might opt for a feature configuration that is unlikely to be realistic~\cite{rudin2019stop}. Thus,  our tool should allow non-expert users to evaluate not only the value of the outcome, but also how realistic the input configuration is when compared to existing data points and configurations.

%% file: sections/6-implementation.tex
\section{Visual Interface}

\sysname{} is a web-based interface. We used Python as the back-end language and D3~\cite{bostock2011d3} for interactive visualization. We used Tetrad~\footnote{\href{https://www.ccd.pitt.edu/tools/}{https://www.ccd.pitt.edu/tools/}}, and semopy~\footnote{\href{https://semopy.com}{https://semopy.com}} for causal analysis.

As per \textbf{DG1}, \sysname{} has two interactive modules. While the Interpretation Module is accessible to both expert and non-expert users, the Model Creation Module is only available to the expert users (\textbf{DG1}). We describe the visual components of these two modules below.

\subsection{Model Creation Module}
The Model Creation Module is divided into five regions (Figure~\ref{fig:create_model}). The Control Panel (A) allows an expert user to upload data and run statistical models on the data. The Central Panel (B) visualizes the causal model obtained from automated algorithms. An expert user can select the appropriate algorithm from a dropdown list there. The graph returned by the automated algorithms is not necessarily a DAG; it can contain undirected edges. Besides, an expert user can edit the graph in this module (\textbf{DG2}), often resulting in a change of the structure of the graph. Since the structure of the graph is uncertain, we decided to use GraphViz~\cite{ellson2001graphviz}, a well-known library for graph visualization. The panel facilitates four sets of edge editing features: (1) Add, (2) Direct, (3) Remove, and (4) Reverse. When editing the causal model, an expert user can evaluate several model fit measures in panel C, model parameters in D, and prediction accuracy in E.

\subsection{Model Interpretation Module}

The interpretation module (Figure 4) uses a different visual representation to present the graph than the model creation module since a parameterized causal model has a definitive structure (DAG). The interpretation module accepts a DAG as input and employs topological sort to present that DAG in a left to right manner.
 
Each variable in the causal model contains two circular knobs: a green and an orange knob. A user can control two different profiles independently by setting the green and orange knobs to specific values (\textbf{DG3}). This two profile mechanism facilitates instance comparison and what-if analysis (\textbf{DG3}, \textbf{DG4}, see Section 7).
 The range for the input knobs is set from the min to the max of a particular variable. Each knob provides a grey handle which a user can use to move the knob through mouse drag action. The user can also set the numbers directly in the input boxes, either an exact number or even a number that is out of range (outside $(max-min)$ range) for that variable. In case of an out of range value, the circular knob is simply set to min or max, whichever extrema are closer to the value. 
 The outcome variable is presented as a bar chart in the causal model. Similar to the input knobs, the outcome variable contains two bars to show prediction values for two profiles.  
 

Finally, we follow the visual design of Wang et al. \cite{wang2017visual, wang2015visual} to encode the edge weights in the causal DAG. To visualize intervention, all edges leading to an endogenous variable are blurred whenever an user sets that variable to a specific value. Consequently, the user can cancel out the intervention by clicking the $\times$ icon beside an endogenous variable in which case its value is estimated from its parent nodes and the edges return to their original opacity.



\subsubsection{Profile Map}
The profile map (Figure 4(C)) is a biplot which shows the nearest neighbors of a profile (\textbf{DG4}). To compute the biplot, we run PCA on the selected points. A user can control the radius of the neighborhood, given by the range of the outcome variable, through the ``Radius'' slider. The neighbors are colored according to the outcome value, as specified in the color map on the right of the plot. 

A user can hover the mouse over any point on the map to compare the data point with the existing green profile (\textbf{DG4}). Subsequently, the user can click on any point to set that point as a comparison case for a more detailed analysis. Both green and orange disks (larger circles) move around the map as the user changes the profiles in the causal model.


\subsubsection{Evolution Tracker}
One of the fundamental features of any UI is the support for a redo and undo operations.  We introduced the Evolution Tracker (Figure  4(D)) to facilitate this. The tracker is a simple line chart with two lines for two profiles in the system. The $x$-axis represents the saved state while the $y$-axis shows the outcome value at that particular state. A user can click on the ``Save Profiles'' button to save a particular state in the tracker and can click on any point in the tracker to go back and forth between different states.



\subsubsection{Similarity (or Realism) Meter}
We introduced the similarity (\textit{realism}) meter to allow users to determine how common their profile is compared to that of the existing population captured by the dataset (\textbf{DG5}). It is a safety check for unreasonable expectations that could be generated by the interactive interpretation module, and so it is an important part of the interface. 

To realize it, we opted for a multi-dimensional method similar to detecting an outlier in one dimension using the $z$-score. At first, we fit a Gaussian Mixture Model on the existing data points in multivariate space. A mixture model with $K$ Gaussians or components is defined as:
\begin{equation}
\begin{split}
    P(X) & =  \prod_{n=1}^N \sum_{k=1}^K P(X_n |C_k) P(C_k) 
     = \prod_{n=1}^N \sum_{k=1}^K \phi_k N(X_n | \mu_k, \Sigma_k )
\end{split}
\end{equation}
where $N$ is the number of datapoints, $\phi_k =  P(C_k)$ is the mixture weight or prior for component $k$, and $\mu_k, \Sigma_k$ are the parameters for the $k$-th Gaussian. Once the parameters are learned through the Expectation-Maximization algorithm, we can calculate the probability of a datapoint $x$ belonging to a component $C_i$ using the following equation
\begin{equation}
    \begin{split}
        P(C_i|x) & =  \frac{P(C_i)P(x|C_i)}{\sum_{k=1}^K P(C_k)P(x|C_k)}  = \frac{\phi_i N(x| \mu_i,\Sigma_i)}{\sum_{i=1}^K \phi_k N(x| \mu_k,\Sigma_k)}
    \end{split}
\end{equation}

A high value of $P(C_i|x)$ implies that $x$ is highly likely to belong to $C_i$, whereas a low value $P(C_i|x)$ implies that the features of $x$ is not common among the members of $C_i$. Thus, $P(C_i|x)$ can be interpreted as a scale of how ``real'' a datapoint is to the other members of a component. We translate $P(C_i|x)$ to a human understandable meter with $P(C_i|x) = 0$ interpreted as ``Rare'', $P(C_i|x) = 0.5$ as ``Moderately Common'', and $P(C_i|x) = 1$ as ``Common''. 


%% file: sections/5-use_cases.tex
\section{Usage Scenario}

In this section, we present a usage scenario to demonstrate how a hypothetical expert user (Adam), and a non-expert user (Emily) could benefit from \sysname{}. 

Adam (he/him) is a Research Engineer at a technology company and is responsible for creating a housing price prediction model. Non-expert users will eventually use the model. As a result, Adam also needs to create an easy-to-understand interactive interface for the non expert users. Based on these requirements, Adam decides to use an interpretable model for prediction and determines that \sysname{} matches the requirements perfectly.


\subsection{Creating the Model}
Adam starts off \sysname{} by uploading the housing dataset~\cite{harrison1978hedonic} into the Model Creation Module (Figure~\ref{fig:create_model}).  Next, Adam selects the PC algorithm~\cite{glymour2019review} for searching the causal structure (not depicted). Upon seeing the causal DAG obtained from the PC algorithm, Adam uses prior knowledge to refine the causal relations. For example, Adam notices that the initial model has an undirected edge between ``INDUSTRIALIZATION'' and ``DISTANCE\_FROM\_CITY''. From domain expertise, Adam knows that ``DISTANCE\_FROM\_CITY'' can be a cause of ``INDUSTRIALIZATION'', but the opposite relation is not plausible. Adam directs the edge from ``DISTANCE\_FROM\_CITY'' to ``INDUSTRIALIZATION'' and notices that the model fit measures (Figure~\ref{fig:create_model}(C)) have also increased. Figure~\ref{fig:create_model} presents the final causal DAG obtained in this iterative process (see supplemental video for the intermediate steps). The final model fit measures are: Comparative Fit Index (CFI): 0.951; Goodness of Fit (GFI): 0.950; Adjusted Goodness of Fit (AGFI): 0.919; and RMSEA: 0.0997.  The measures indicate a good SEM model fit. 
Satisfied by the model performance, Adam hits the ``Save Model'' button and moves to the ``Model Interpretation'' module.



\subsection{Interacting with the Causal DAG and Exploring Nearest Neighbors}
After creating the model, Adam wants to explore and verify the model in the Interpretation Module (Figure 4) before making it public. Adam starts off the exploration process by selecting a sample data row from Figure~\ref{fig:explain_model}(B). Adam observes that the selected row is immediately reflected in the feature values and the outcome of the model has changed to 22.61K. Adam also observes that the edges entering the endogenous variables became blurred (deactivated) since those variables were set to reflect the selected row and can no longer be estimated from the exogenous variables. From the profile map in Figure~\ref{fig:explain_model}(C), Adam notices that the selected housing has a relatively small price. Adam decides to compare the selected housing with higher prices. To do so, Adam selects a data point with a higher housing price from the profile map. Immediately, an orange profile is created in the causal DAG. From the two profiles, Adam easily understands where the two housing differed and how that affected the outcome. At this point, Adam hits the ``Save Profile'' button to save both the profiles in the tracker (Figure~\ref{fig:explain_model}(D)). In a similar manner, Adam explores several other data points to get a concrete idea of the model. At the end of the analysis, Adam is confident that the model is accurate, interpretable, and is ready for deployment. Adam then publishes the interface with the name \textit{housingX}.

\begin{figure}
    \centering
    \includegraphics[width=0.95\columnwidth]{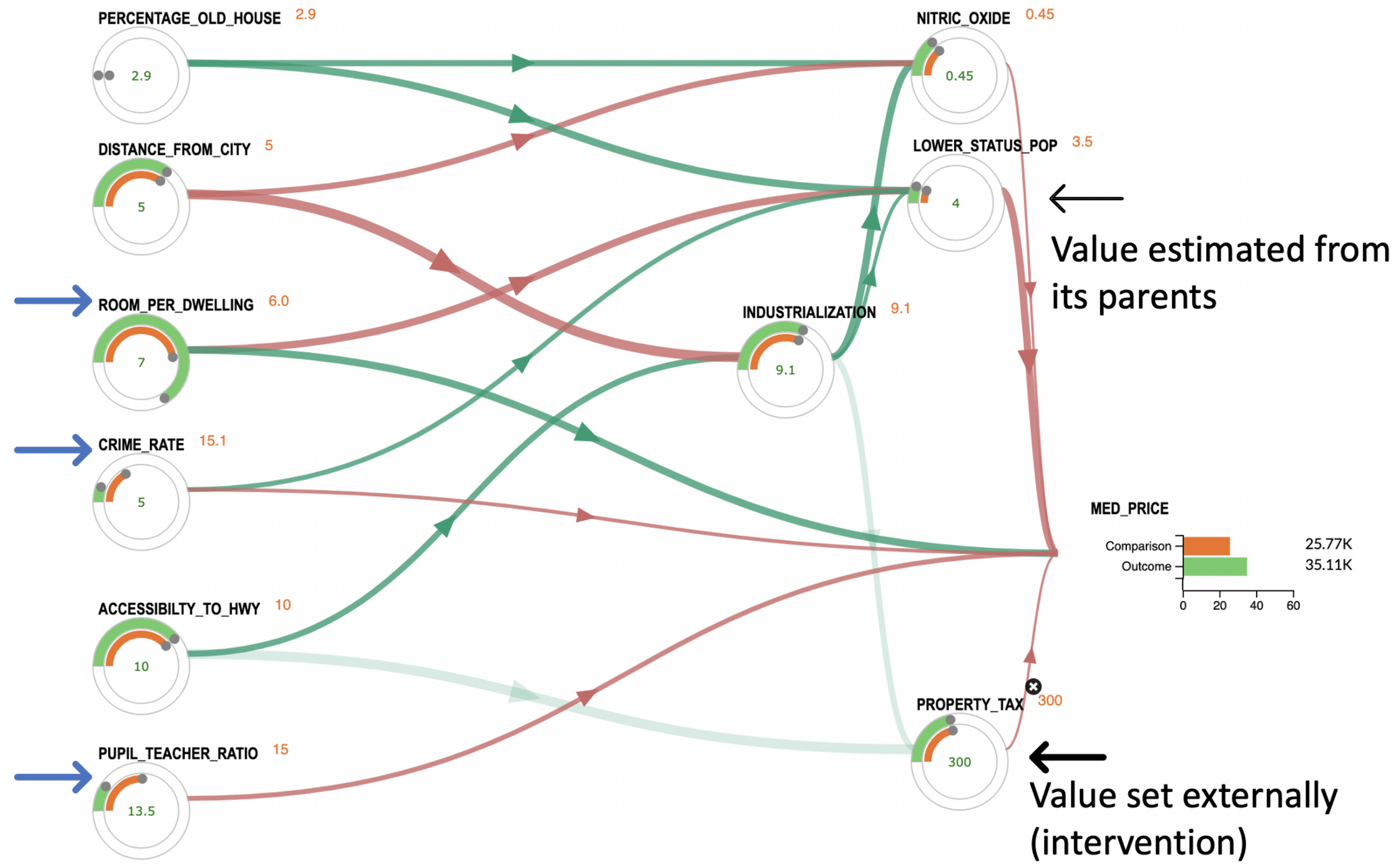}
    \caption{Asking what-if questions in \sysname{}. A user can keep one profile (green) fixed, and change the other profile (orange) to ask what-if questions. The blue arrows indicate the changes in the orange profile. Note that property tax is set to 300 by the user. As a result, changing its parents will not affect property tax. The other endogenous variables are estimated from their parents.}
    \label{fig:my_label}
\end{figure}

\subsection{Understanding the Causal Relations}

Emily is a middle-aged female (she/her) who would like to purchase a new house. Emily decides to understand  the quality of the desired neighborhood through \textit{housingX}. 


Emily visits the site, which features the Interpretation Module of Outcome-Explorer, and starts off by watching a small tutorial on how to use the interface. After that, Emily starts to make sense of the variables and how they are connected to each other. Emily notices that Property Tax is calculated from a region's accessibility to the highway and the industrialization index. Emily further notices that Industrialization depends on both the region's distance from the city and its accessibility to the highway. Based on that, Emily concludes that Property tax depends on three factors: Accessibility to Highway, Distance from City, and Industrialization.
Interestingly, Emily observes that the Median Price has a red edge from Property Tax, meaning houses located in areas with higher property taxes are priced lower than houses from areas with lower property taxes. 

\subsection{Exploring Outcomes \& What-if Analysis}
After understanding the causal relations between variables, Emily now starts putting values to the knobs in the interface. Emily observes that once the value of an internal node is set, edges leading to that node become blurred. For example, Emily sets the value of Property Tax to 300, which decreases the Median Price of the houses, but also blurs the edges entering Property Tax (Figure~\ref{fig:my_label}). 


After finding the ideal neighborhood, Emily notices that the Median Price of that neighborhood is around \$35,000. But, Emily only has a budget of around \$25,000. Based on that, Emily decides to change the variables so that the median price comes down to \$25,000. Emily fires off the comparison mode by clicking the ``Initialize Comparison'' Button. Immediately, a new orange profile is created in the interface which is exactly the same as the current green profile. Keeping the green profile constant, Emily iteratively changes the variables to take down the median housing prices to \$25,000. In this iterative process, Emily utilizes the tracker regularly to go back and forth between different profile configurations. Emily also consults the realism meter regularly to see how common the selected housing is compared to the existing neighborhoods. She confirms that the orange line stays on the right end of the meter which means that these configurations are very common (Figure~\ref{fig:explain_model}).



%% file: sections/7-evaluation.tex
\section{Evaluation}
We evaluated \sysname{} in two phases: (1) we conducted think-aloud sessions with three ML practitioners to gather expert feedback and gauge the system's real-world potential; 
(2) we conducted a user study with $18$ users to assess the effectiveness and usability of \sysname{} in supporting the explanation needs of non-expert users.

\subsection{Expert Evaluation}
We invited three ML practitioners as expert users (1 female, 2 male) to examine \sysname{}. Participation was voluntary with no compensation. All three participants had post-graduate degrees and had conducted research in the field of XAI, Fairness, and Data Ethics for at least five years. They were also familiar with statistical causal analysis.  

The tool was deployed on a web server and the sessions were conducted via Skype. Participants shared their screen as they performed the tasks. One author communicated with the participants during the sessions while another author took notes. Each session started with the participant observing a live demo of the tool. After that, participants were asked to choose one out of two datasets: Boston Housing~\cite{harrison1978hedonic} and the PIMA Diabetes dataset~\cite{smith1988using}. Once a participant chose a dataset, we provided them with
the textual descriptions of the features and a task list. The task list was designed to guide the participants in exploring different components of \sysname{}. Participants started off by creating a causal model using the Creation Module, and then gradually moved into examining different explanation methods available in the Interpretation Module.  While performing the tasks, participants thought-aloud and conversed with the authors continuously. We sorted their feedback in the four thematic categories, as described next. 


\subsubsection{Comprehensive and Generalizable}
All three expert users found the ``Model Creation'' module to be ``comprehensive'', and ``generalizable''. Participants found the accuracy statistics to be most helpful as that feature is not available on other comparable causal analysis tools. According to E1: \sysname{} was ``rigorous'' in terms of causal functionality and should enable users to obtain the ``best possible causal model''.

The participants also found the interpretation module to be comprehensive. E1 mentioned that the interactive causal DAG alone should allow non-experts to understand the model. Additionally, they found the two profile comparison mechanism to be helpful and appreciated the fact that the user can ask the what-if questions directly to the model.

\subsubsection{Engaging, Thought Provoking, and Fun}
Participants continuously engaged themselves in making sense of the causal relations. Throughout the session, they enthusiastically initiated discussions with the authors to share their personal experiences related to causal relations.

Participants also found the visual design of the Interpretation Module to be ascetically pleasing and fun to interact with. They mentioned that the interface has a ``certain gaming flavor'' to it. E3 opined that the thought-provoking and interactive nature of \sysname{} might entice curious non-expert users to gather knowledge on a domain of interest. 




\subsubsection{Prior Knowledge and Position in the ML Pipeline}

Participants suggested that \sysname{} could be used once an expert user has preprocessed and explored the dataset. It would provide users the necessary background knowledge for creating and explaining the causal model. According to E1,

\textit{``I can see that the user might need to tweak the initial causal model iteratively to reach the final model, but that is also true for many ML models. The process of creating a predictive model is often messy, and requires several iterations, each of which requires users to utilize prior knowledge to refine the model.''}








%


\subsubsection{Disclaimers} 
Causal relations make stronger claims than associative (correlative) relations. Participants suggested that the implication of the causal relations should appropriately be communicated to the end-users. For example, a particular causal relation may hold true for a particular task or domain, but not in general. Expert users should be aware of such potential misleading relations in the causal model, and should provide disclaimers to the non-expert users whenever needed. This will ensure that non-experts are not misled into thinking that the causal relations in \sysname{} are ubiquitously true.


\subsection{User Study with Non-expert Users}

 To understand how non-expert users might benefit from \sysname{} we conducted a user study. We aimed at validating the following hypotheses:
\begin{itemize}
    \item \textbf{H1}: \sysname{} will improve a user's understanding of the embedded predictive causal model in comparison to the state-of-the-art explanation method. 
    \item \textbf{H2}: \sysname{} will increase a user's efficiency in reaching a desired outcome in comparison to the state-of-the-art explanation method. 
    \item \textbf{H3}: \sysname{} will be easy to use.
\end{itemize}

 We chose SHAP~\cite{NIPS2017_7062} as a comparison case for \sysname{} as it is a prevalent and widely used post-hoc explanation technique. 
 Another motivation for comparing our approach with SHAP was the interpretable nature of our tool. SHAP approximates the prediction mechanism of a model without showing the model itself, an approach fundamentally different than our.
Additionally, SHAP is open-source and provides several visualizations to aid the explanations, ensuring a fair comparison with our visual interface. 
Hence, we conducted a repeated-measures within-subject experiment with the following two conditions.

\begin{itemize}
    \item[\textbf{C1.}] \textbf{SHAP}: This condition included input boxes which users could use to change variables. Users had access to two charts provided by SHAP: a bar chart showing global feature importance and a variant of stacked bars (force plot) showing the feature contribution for a decision (see supplemental material).
   
    \item[\textbf{C2.}] \textbf{\sysname{}-Lite}: This prototype included only the interactive causal DAG of the Interpretation Module with other components hidden (see supplemental material). 
\end{itemize}


 We chose to include only the causal DAG in the study as the other components provide auxiliary tools to understand the model, but are not necessary to interpret the model. The inclusion of these components could hinder a fair comparison between \sysname{} and SHAP. To minimize the learning effect, we included two datasets and counterbalanced the ordering of study conditions and datasets. These datasets are the Boston housing and the PIMA Diabetes dataset, both of which appeared previously in the XAI literature~\cite{hohman2019gamut,ming2018rulematrix}. 

\subsubsection{Participants}
We recruited $18$ participants (10 males, 8 females) through local mailing lists, university mailing lists, and social media posts.  
Participation was voluntary with no compensation.  
The participants varied in age from $19$ to $35$ ($M=25$, $SD=4.2$1). None of the participants had machine learning expertise. The participants were comfortable in using web technology and had a high-level idea of automated decision-making through exposure to credit-card approval and loan approval systems. Additionally, two participants had experience with interactive visualization through interactive online news. All participants reported a basic understanding of the dataset domains (housing and diabetes) in the post-study interview, but did not report any specific expertise on the domains.

 
 \begin{figure*}
     \centering
     \subfloat[Average number of changes and magnitude of changes on non-impacting variables]{
        \includegraphics[width=0.30\textwidth,height=0.25
        \textwidth]{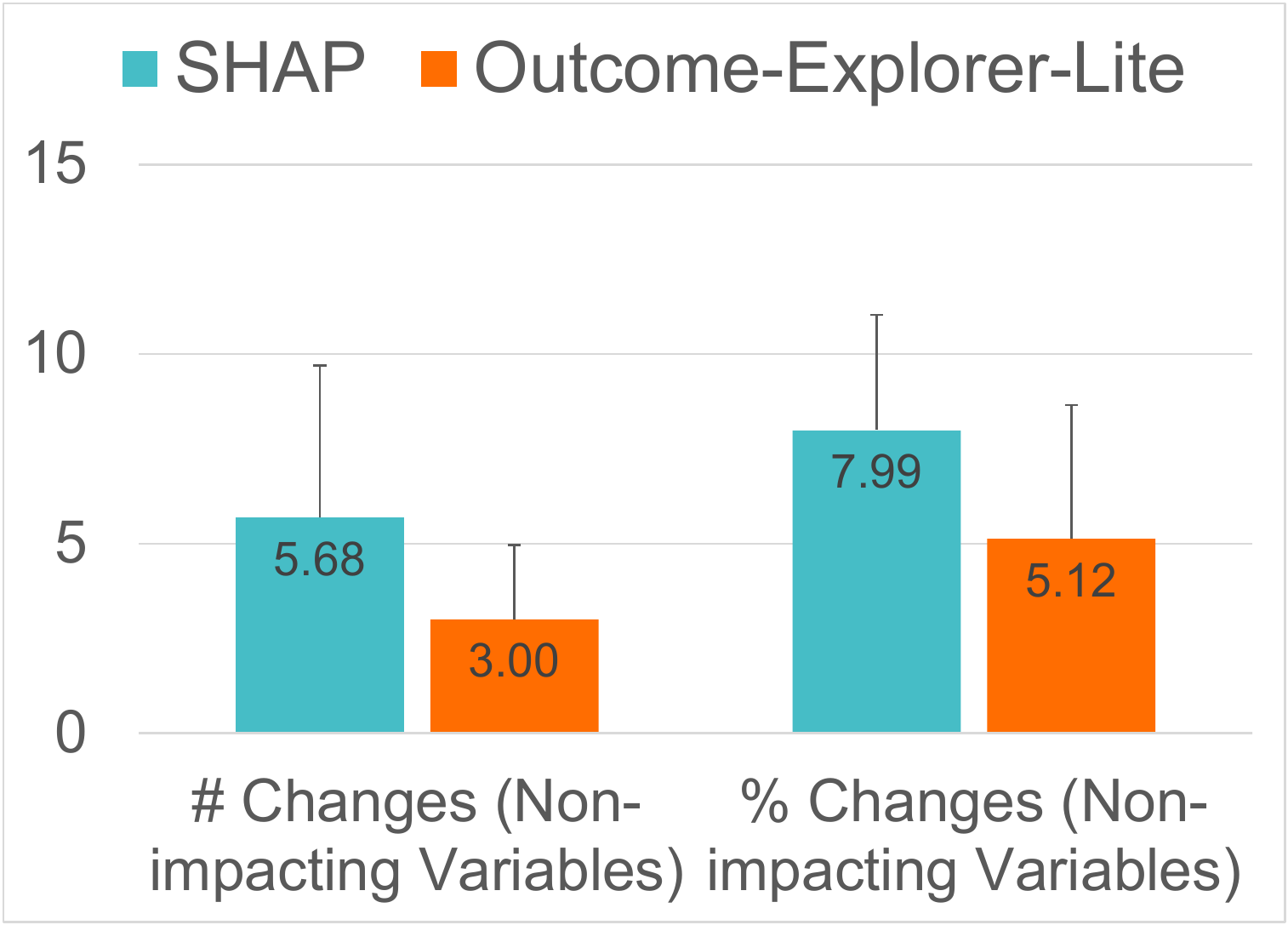}
        \label{fig:quantitative}
     
     }
     \quad
     \subfloat[Average number of changes and magnitude of changes on all variables]{
        \includegraphics[width=0.30\textwidth,height=0.25
        \textwidth]{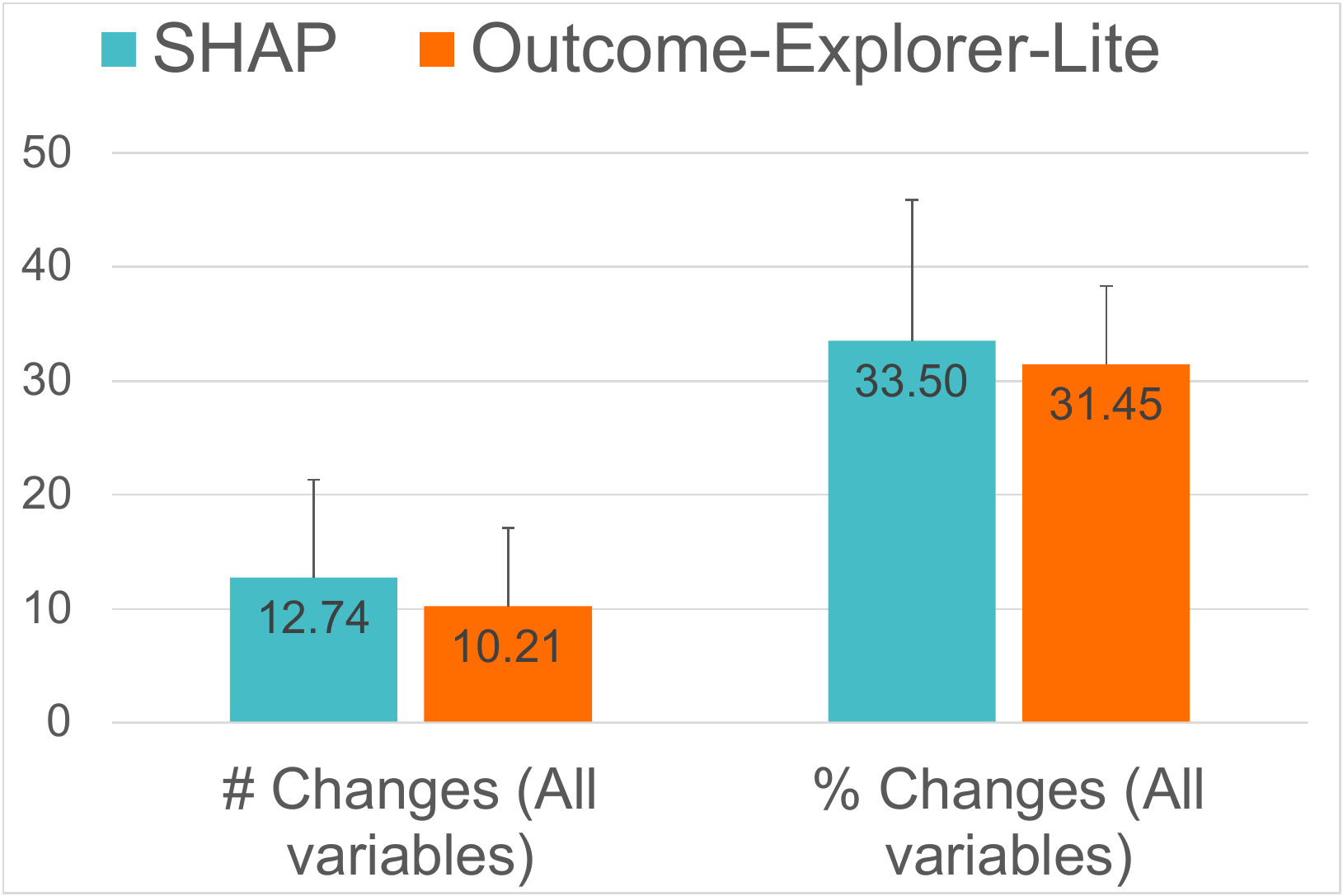}
        \label{fig:quantitative1}
     
     }
     \quad
     \subfloat[Subjective Measures (1: Strongly Disagree, 7:Strongly Agree).]{
        \includegraphics[width=0.33\textwidth,height=0.25
        \textwidth]{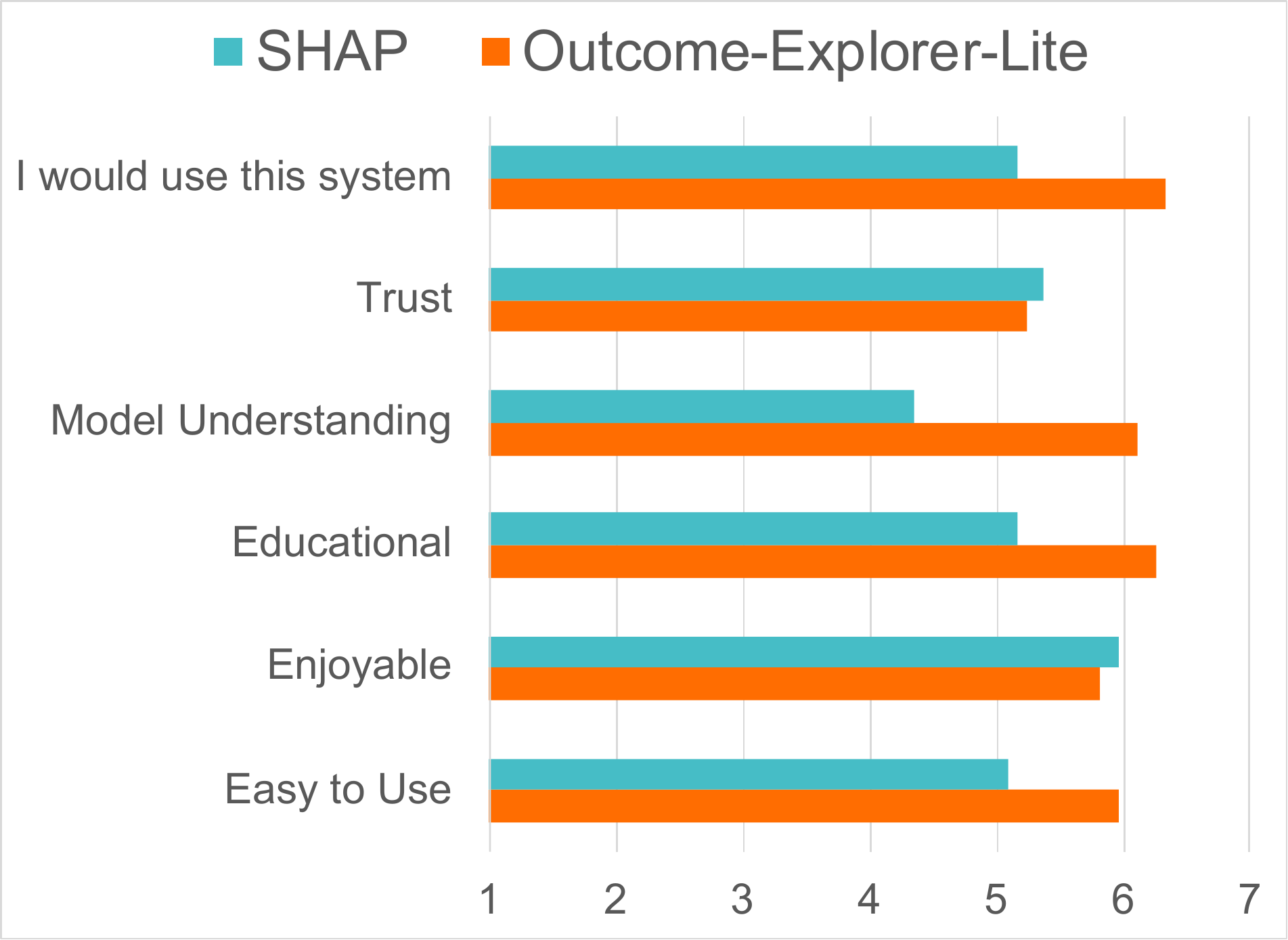}
        
        \label{fig:subjective}
    }
      \caption{Study Results. The average number of changes and the average magnitude of changes (\%)  made to (a) non-impacting variables, and (b) all variables to reach the target outcomes. (c) Average self-reported subjective measures. Error bars show $+1$ SD.}
     \label{fig:self-reported}
 \end{figure*}

\subsubsection{Tasks}
XAI interfaces are frequently evaluated on ``proxy tasks'' such as how well humans predict the AI’s decision, and subjective measures of trust and understanding~\cite{buccinca2020proxy}. Recent research suggests that proxy tasks and subjective measures are not good predictors of how humans perform on actual decision-making tasks~\cite{buccinca2020proxy}. Based on that, we decided to evaluate our tool on actual decision-making tasks.
The tasks were similar to the case study presented in Section 7.4. For example, in the case of the housing dataset, we provided the participants with the scenario of a person who wants to buy an ideal housing (e.g. housing with price 35K) with budget constraints. We then asked the participants to reach alternative/target outcomes (e.g. reducing housing price from 35K to 25K) to satisfy the budget constraints while minimizing the number of changes, and the magnitude of the changes from the ideal housing.
  Note that both conditions made predictions based on the same underlying causal model. We also collected self-reported subjective measures such as model understanding, trust, and usability.
  
  
 
 
 
\subsubsection{Study Design}



 Similar to the sessions with the expert users, we conducted the study sessions via web and Skype. A study session began with the participant signing a consent form.  Following this, the participants were introduced to the assigned first condition and received a brief description of the interface. The participants then interacted with the system (with a training dataset), during which they were encouraged to ask questions until they were comfortable. Each participant was then given a scenario and a task list for the first condition. After completing the tasks, participants rated the study conditions (interfaces) on a Likert scale ranging from $1$ (Strongly Disagree) to $7$ (Strongly Agree) based on six subjective measures. The same process was carried out for the second condition. Each session lasted around $\sim$1 hour and ended with an exit-interview.

 \subsubsection{Results}


 \textbf{H1: Model Understanding.}
 In a causal model, the exogenous variables may not affect the outcome if endogenous variables are set to specific values. While \sysname{} visualizes this interplay, SHAP only estimates feature contributions to the decision and it does not explain why some variables are not affecting the outcome. We refer to such variables as \textit{non-impacting variables}. We anticipated that interactions with non-impacting variables might reveal how well users understood the model. Based on that, to account for \textit{model understanding}, we measured (1) \textit{the number of changes (non-impacting)}  and (2) \textit{the magnitude of changes (\%, non-impacting)}. Here, non-impacting refers to the changes made on non-impacting variables. We used a paired t-test with Bonferroni correction and Mann-Whitney U to assess statistical significance of the quantitative and likert scale measures respectively. 
 

  On average, the participants made $5.68$ ($SD=4.01$) changes to the \textit{non-impacting} variables when using SHAP  
 while for \sysname{}-Lite the average was $3.00$ ($SD=1.97$).  Participants  reduced the changes made to the non-impacting variables by $47\%$, which was statistically significant ($p<0.02$); Cohen's effect size value ($d=0.68$) suggested a medium significance. We also  found a significant difference between the magnitude of changes users made on non-impacting variables ($36\%$ reduction with $p<0.001$, plotted in Figure~\ref{fig:quantitative}). 
 
 In order to understand how study conditions and dataset relate to each other with respect to the above quantitative measures, we constructed two mixed-effect linear models, one for each measure. We tested for interaction between study condition and dataset while predicting a specific measure. While there were no interaction effects and dataset did not play any significant role in predicting the measures, we found study condition to be the main effect in predicting the number of changes on non-impacting variables ($F(1,26.033)=7.723,p=0.01$) and the magnitude of changes (\%)  on non-impacting variables ($F(1,26.992)=13.140,p=0.001$).
 
 Finally,  as shown in Figure~\ref{fig:self-reported}(c), participants rated \sysname{}-Lite favorably in terms of Model Understanding ($M:6.11, SD:0.50$), and   Educational ($M:6.26,SD:0.87$). In comparison, the scores for SHAP were: Model Understanding ($M:4.34, SD:0.911$), and Educational ($M:5.15,SD:0.964$). The differences were statistically significant with $p< 0.0001$ (Model Understanding) and $p<0.01$ (Educational). However, we also observe that participants' trust did not improve in \sysname{}. In the post-study interview, several participants mentioned their unpleasant experiences with automated systems. Such distrusts are unlikely to be changed in one study, and that might be the reason for equal trust in both conditions.

 \textbf{H2: Efficiency in Decision-making Tasks.}
 We measured two quantitative measures to account for users' \textit{overall performance} in decision-making tasks. They are the total (1) \textit{number of changes}, and (2) \textit{magnitude of changes (\%)} made to input variables. As shown in Figure~\ref{fig:quantitative1}, the average number of changes were $12.74$ ($SD=8.59$) for SHAP, and $10.21$ ($SD=6.88$) for \sysname{}-Lite.  The difference was not statistically significant.  We also did not find a significant difference between the overall magnitude of changes (\%) users made in each study condition. Finally, we measured the time taken to complete the tasks, but no statistically significant difference was found.
 
Similar to the above, we constructed two mixed-effect linear models. We did not find any interaction effects, and the datasets as well as the conditions did not play any significant role in predicting the measures.

 \textbf{H3: Ease of Use.} 
Participants rated \sysname{}-Lite favorably in terms of Easy to use ($M:5.96,SD:0.96$), and I would use this system ($M:6.33,SD:0.6$). In comparison, the scores for SHAP were: Easy to use ($M:5.08,SD:1.35$),  and I would use this system ($M:5.16,SD:1.42$).
The differences were statistically significant with $p< 0.04$ (Easy to use) and $p<0.03$ (I would use this system). The other metric (Enjoyable) was not statistically significant.




The results matched our anticipation that \sysname{}-Lite will improve user model understanding and that they will learn more about the prediction mechanism using our tool. In the post-study interview, participants appreciated the visual design of the causal DAG which might be the reason why they found \sysname{}-Lite to be easy to use and want to see it in practice. 

On average, participants spent slightly more time when using \sysname{}. While familiarizing with the interface was one factor for that, in the post-study interview, several participants mentioned that they felt curious, spent more time to learn the relations, and put some thought before taking an action. A participant, a senior college student, mentioned:
\textit{``The interface (\sysname{}-Lite) is fun, attractive as well as educational. I feel like I learned something.  I did not know much about housing prices before this session. But, I think I now have a much better understanding of housing prices. If available in public when I buy a house in the future, it will help me make an informed decision.''}

%% file: sections/8-discussion_conclusion.tex
\section{Discussion and Limitations}


\textbf{Model Understanding vs Overall Performance:} 
The user study validated H1 and H3, but not H2. 
 The study revealed that participants reduced interactions with non-impacting variables significantly in \sysname{}-Lite, indicating a better model understanding compared to SHAP. The lack of edges or blurred edges in \sysname{}-Lite provided participants with clear evidence for non-impacting variables. On the other hand, while using SHAP, participants interacted with the non-impacting variables despite observing their zero feature importance in the visualization. They constructed several hypotheses about non-impacting variables while using SHAP, including the possibility of a change of impact in the future, and their indirect effects on other variables. This may be the reason for the increased interaction with the non-impacting variables. However, improved understanding of non-impacting variables did not result in better overall performance. Participants instead increased interaction with the impacting variables to understand the effects of causal relations while using \sysname{}-Lite. As a result, the overall performance (i.e., the total number of interactions with impacting and non-impacting variables) remained similar for both conditions. 
 We also believe that the increased focus on impacting variables while using \sysname{}-Lite fostered the observed better model understanding in the subjective measures. 
 

%





 
 \textbf{The Accuracy-Interpretability Trade-off:}
 We acknowledge that causal models might not reach the prediction accuracy of complex machine learning models. The comparison requires rigorous experiments on common ML tasks which is beyond the scope of this paper. However, there exists empirical evidences where causal models or linear models such as ours outperformed complex ML models~\cite{rudin2019we,tople2019alleviating}. As stated by Rudin~\cite{rudin2019stop}, the idea that interpretable models do not perform as well as black-box models (the \emph{accuracy-interpretability tradeoff}) is often due to the lack of feature engineering while building interpretable models.

\textbf{Implication for XAI Research: }
\sysname{} offers several design insights for visual analytics systems in XAI. First, its novel two module design shows that it is possible to support the explanation needs of experts and non-experts as well as the model creation functionalities for experts in a single system. While we acknowledge this is not a strict requirement for an XAI interface, we believe our work will motivate non-expert inclusive design of XAI interfaces in the future.
Second, it shows that an effective XAI interface does not necessarily require a new and complex visualization. ``Simple'' visual design and ``intuitive'' interactions are effective ways to convey the inner workings of predictive models, especially for non-expert users.

 
 
 Another potential impact of \sysname{} is bridging XAI and algorithmic fairness research. Algorithmic fairness ensures fair machine-generated decisions while XAI ensures transparency and explainability of ML models. Although highly relevant, these two research directions have not been bridged together yet. By promoting the causal model, a highly effective paradigm for bias mitigation strategies, \sysname{} opens the door for an ML model to be transparent, accountable, and fair altogether.
 
 Finally,  
 our design and visual encoding can be extended to other graphical models. For example, a Bayesian Network is also represented as a DAG, and the design of \sysname{} can be transferred to interactive XAI systems based on Bayesian networks.
 

\textbf{Limitations \& Future Work:}
It is important for a causal model to have a sufficient number of variables that cover all or at least most aspects determining the predicted outcome.  Causal inferencing under incomplete domain coverage can result in islands of variables or a causal skeleton where some links are reduced to correlations only. We are currently experimenting with evolutionary and confirmatory factor analysis to introduce additional variables that can complement the native set of variables. These variables are often not directly measurable and can serve as latent variables. Our preliminary work has shown that they can greatly add to both model comprehensibility and completeness. 

Another source of error can be confounders which can lead to an overestimation or underestimation of the strength of certain causal edges. There are several algorithms available for the detection and elimination of counfonding effects and we are presently working on a visual interface where expert users can take an active role in this type of effort. 

A current limitation of our system is scalability. At the moment we limit the number of variables to George Miller’s Magical Number Seven, Plus or Minus Two paradigm~\cite{miller1994magical}. This allowed us to understand the explanation needs of mainstream non-expert users  and supporting them through interactive visualizations on previously studied XAI datasets~\cite{hohman2019gamut,ming2018rulematrix}. However, in many real-life scenarios, there can be ``hundreds of variables'', which could overwhelm non-expert users. 
We envision that the Model Creation Module could be enhanced with advanced  feature engineering capabilities, such as clustering, dimension reduction, pooling of variables into latent variables (factor analysis), and level of detail visualization~\cite{zhang2012network}. Alternatively, scalable causal graph visualization~\cite{xie2020visual} could also be used for this purpose. Our future work will focus on gaining more insight on how much complexity non-expert users can handle, and which of the above-mentioned methods work best for them.

Finally, so far the participants we have studied were all from the younger generation (19-35) who generally tend to be savvier when it comes to the graphical tools used in our interface. In future work we aim to study how our system would be received by older members of society. It might require additional information integrated into the user interface, such as tooltips and pop-up suggestion boxes and the like. 











\section{Conclusion}

We presented \sysname{}--- an interactive visual interface that exploits the explanatory power of the causal model and provides a visual design that can be extended to other graphical models. \sysname{} advances research towards interpretable interfaces and provides critical findings through user study and expert evaluation. 

We envision a myriad of applications of our interface. For example, bank advisors or insurance agents might sit with a client and use our system to discuss the various options with them (in response to their right to explanation), or a bank or insurance would make our interface available on their website, along with a short instructional video.
Future work will explore how complex a model can get while still being understandable by non-expert users.  